\documentclass[12pt,a4paper]{article}
\usepackage{hyperref}
\usepackage{amsmath}
\usepackage{fullpage}
\usepackage[english]{babel}
\usepackage{titlesec}
\usepackage{graphicx}
\usepackage{physics}
\usepackage{amsfonts}
\usepackage{subfig}
\usepackage{float}
\usepackage{epsfig}
\usepackage{setspace}
\usepackage{authblk}
\usepackage{cite}
\usepackage{xcolor}
\DeclareMathOperator{\e}{e}

\begin{document}

\title{Mirrors-light-atoms entanglement in ring optomechanical cavity}

\author{Oumayma El Bir \thanks{oumayma.elbir@um5s.net.ma}}
\author{Morad El Baz \thanks{morad.elbaz@um5.ac.ma} }
\affil{ Mohammed V University of Rabat, Faculty of Sciences, ESMaR, Rabat, Morocco}

\date{\today}
\maketitle

\begin{abstract}
The present paper illustrates the realization of an atom-optomechanical system where an atomic ensemble is confined in a ring optomechanical cavity consisting of a fixed mirror and two movable ones. An analysis of the dynamics and the  linearization of the equations allows to derive the multimode covariance matrix. Under realistic experimental conditions, we numerically simulate the steady-state bipartite and tripartite continuous variable entanglement using the logarithmic negativity, and analyze the shared entanglement in the multimode system. The introduction of the atomic medium allows to obtain a larger plateau for the entanglement and make more resilient to the temperature decohering effects.

\end{abstract}

\section{Introduction}
The field of cavity optomechanics have witnessed a rapidly growing advance with the goal of controlling the interaction between electromagnetic radiation and nanomechanical motion \cite{1,2,3,4,5,6}. These systems, can be coherently manipulated by pumping the cavity with an external laser field which leads to an optomechanical coupling between the cavity mode and the mechanical oscillator. In this context, most experimental and theoretical efforts have shown that hybrid optomechanical systems can be used to entangling mechanical resonators, atoms, and optical cavity fields in different ways. For instance, entangling two mirrors of two different cavities driven by entangled light beams was proposed in \cite{7}. One can also achieve an entanglement between two movable mirrors in an optomechanical cavity as in \cite{8}, entangling a nanomechanical oscillator with a Cooper-pair box in \cite{9}, entangling two mirrors in a ring cavity by using a phase-sensitive feedback loop \cite{10} and atomic ensembles \cite{11, 12, 13, 14, 15, 16}. Enhancing quantum correlations and entanglement \cite{18} can be achieved via cross- Kerr nonlinearity \cite{17}. 

Recently, research on the possibilities of constructing entangled states between the macroscopic and subatomic scales in coupled optomechanical systems have been exploited, where an atomic ensemble is placed inside a cavity with a moving mechanical object, which represents a novel type of hybrid systems. These atomic cavity optomechanical systems avoid the difficulties of controlling quantum behavior, decoherence, and quantum-classical boundaries \cite{19}. To explore the impact of the atomic ensemble on the entanglement between mechanical resonators and optical fields, we generate a strongly quantum entanglement between the atomic ensemble and a vibrating mirror by increasing the coupling strength. Another point of interest in this paper is to investigate the genuine tripartite entanglement sharing, since the system describes a four-partite mode (cavity field – atomic ensemble and two vibrating mirrors).

The main purpose of the present paper is to study the behavior of the quantum entanglement present in an atomic ring cavity composed of a ring cavity with a fixed mirror and two movable ones and an atomic ensemble of two-level atoms placed inside the cavity. When both an oscillating mirror and an atomic medium are present in the cavity, it essentially forms a well-fortified environment to generate multipartite genuine entanglement \cite{20, 21, 22, 23}. We focus on enhancing bipartite and tripartite entanglement between the mechanical resonators, atomic ensemble and optical cavity field; we study the influence of the atomic medium on the coherence of the present system.

The remainder of this paper is organized as follows. In section 2, we present the theoretical model of the hybrid optomechanical system and establish the Hamiltonian, obtain the system dynamics, by solving the differential equations of motion and obtain the set of Quantum Langevin Equations of the system. In section 3, we simulate the stationary bipartite and tripartite entanglement by introducing logarithmic negativity and exploit this result to investigate the sharing structure of tripartite entanglement in such states. In section 4, we conclude our results and summarize with some perspectives of the results.

\section{System and model}
\subsection{System}
\begin{figure}[h]
\centering
\includegraphics[scale=0.5]{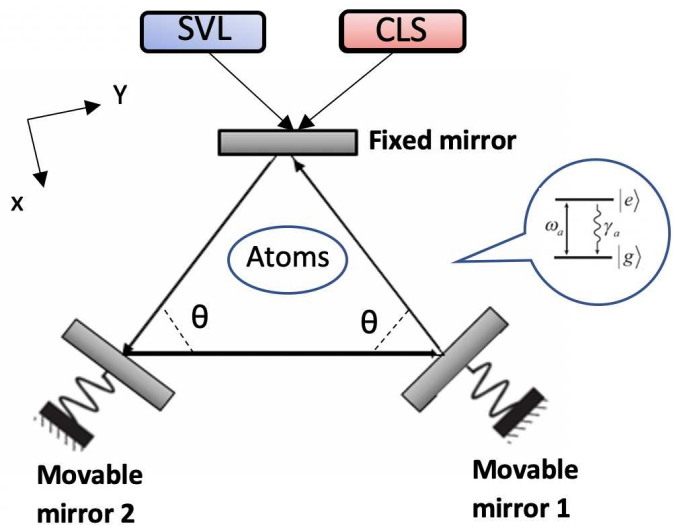}
\caption{Schematic illustration of the Atom-optomechanical system under study including a ring cavity. An ensemble of two-level atoms is placed into the cavity which is coherently pumped by a coherent laser source (CLS) with frequency $\omega_{L}$, and (SVL) a squeezed vacuum light source.} 
\label{ring}
\end{figure}
The model studied here is an Atom-Optomechanical system, made of a ring cavity with length $L$, and an atomic ensemble of two-level atoms placed inside the cavity. This latter is driven by a squeezed light source with frequency $\omega_{S}$ and a coherent laser source with strength $E_{L}$; the setup is schematically shown in Figure \ref{ring}. The optical cavity is composed of two movable mirrors perfectly transmitting and a fixed one partially transmitting in a triangular design. We consider the two movable mirrors as quantum harmonic oscillators with effective masses respectively $m_{1}$, $m_{2}$, and frequencies $\omega_{m1}$ and $\omega_{m2}$, respectively. The cavity field is coupled to the motion of the two mechanical oscillators via the radiation pressure force. The total Hamiltonian of the coupled optomechanical system can be expressed as : $H = H_{0} + H_{interaction} + H_{Drive}$ , which will be well explained in the next section.

This setup is an upgraded version of the ring optomechanical cavity originally introduced in \cite{24}, in which the interaction with an atomic ensemble is considered. It is worth mentioning that systems based on atom-optomechanical models are experimentally feasible \cite{20,25,26}.  
\subsection{The Hamiltonian}

The Hamiltonian of the system is given by $H = H_{0} + H_{Interaction} + H_{Drive}$, with
\begin{eqnarray}
H_{0} &=&  \hbar \omega_{r} a^{\dagger} a + \frac{\hbar \omega_{m1}}{2}(q_{1}^{2} + p_{1}^{2}) + \frac{\hbar \omega_{m2}}{2}(q_{2}^{2} + p_{2}^{2}) + \frac{\hbar \omega_{a}}{2} S_{z}, \\
H_{Interaction}&=& \hbar g (S_{+} a + S_{-} a^{\dagger}) + \hbar \ G_{0} \ a^{\dagger} a  \cos^{2}(\theta /2)(q_{1} - q_{2}),  \\
H_{Drive} &=& i \hbar E_{L} (a^{\dagger} e^{- i \omega_{L}t } - a e^{ i \omega_{L}t }).
\end{eqnarray}

$H_{0}$ represents the free Hamiltonian, the first term of which describes the cavity mode with the resonance frequency $\omega_{r}$, $a^{\dagger}$ and $a$ are the photon  creation and annihilation operators of the optical field, with $[a, a^{\dagger}] = 1$. $q_{j}$ and $p_{j} (j = 1,2)$ describe the dimensionless position and momentum operators of the two mechanical resonators, satisfying $[q_{j}, p_{k}] = i \delta_{jk}$. The last term describes the atomic ensemble composed of  $N_{a}$ two-level atoms, where the frequency $\omega_{a}$ is the transition between the ground state $\ket{g}$ and the excited state $\ket{e}$, having respectively the following energies $E_{g} = - \frac{\hbar \omega_{a}}{2} $ and $E_{e} = \frac{\hbar \omega_{a}}{2} $. $S_{z}$ and $S_{\pm}$ are the spin operators $S_{z, \pm} = \sum_{i= 1}^{N_{a}} \sigma_{z, \pm}^i$, where $\sigma_{z, \pm}^i$ are the Pauli matrices defined by $\sigma_{z}^i = \ket{e}^{(i)(i)}\bra{e} $, $\sigma_{+}^i = \ket{e}^{(i)(i)}\bra{g}$ and $\sigma_{-}^i = \ket{g}^{(i)(i)}\bra{e}$ and satisfying the commutation relations $[\sigma_{+}^{i},\sigma_{-}^{i}] = \sigma_{z}^{i}$ and $[\sigma_{z}^{i},\sigma_{\pm}^{i}] = \pm 2 \sigma_{\pm}^{i}$.   

$H_{Interaction}$ describes the coupling Hamiltonian, where the atoms interact with the two mechanical oscillators via the coupling with the intracavity field, which is called the atom-cavity coupling coefficient and expressed as $g = \mu \sqrt{\frac{\omega_{r}}{2 \hbar \epsilon_{0} V}}$ with $\mu$ being the dipole moment of the atomic transition, $V$ is the volume of the cavity and $\epsilon_{0}$ is the vacuum permittivity and $\hbar$  is Planck constant. The second term takes into account the interaction of the two mechanical resonators motion with the electromagnetic field confined in the cavity due to the radiation pressure force, $G_{0}$ is the optomechanical coefficient,  the angle between the incident and the reflected light on the surfaces of the movable mirrors, $\theta$, (see Figure \ref{ring}).         

The part $H_{Drive}$ represents the drive laser input, where $E_{L} = \sqrt{\frac{\kappa P}{\hbar \omega_{L}}}$, with $P$, $\omega_{L}$ being respectively the power and the frequency of the driven laser, $\kappa$ is the decay rate of the optical cavity.

Working in the rotating frame at the input laser frequency $\omega_{L}$, the Hamiltonian of the system simplifies to   
\begin{equation}
\begin{aligned}
H =& \hbar \Delta_{r} a^{\dagger} a + \frac{\hbar \omega_{m1}}{2}(q_{1}^{2} + p_{1}^{2}) + \frac{\hbar \omega_{m2}}{2}(q_{2}^{2} + p_{2}^{2}) + \hbar \Delta_{a} c^{\dagger} c + \hbar G_{a} (c^{\dagger} a + c a^{\dagger})\\  &+ \hbar G_{0} a^{\dagger} a
 \cos^{2}(\theta /2) (q_{1} - q_{2}) + i \hbar E_{L} (a^{\dagger} - a),
\end{aligned}
\label{hami}
\end{equation}
with $\Delta_{r} = \omega_{r} - \omega_{L}$ and $\Delta_{a} = \omega_{a} - \omega_{L}$ are respectively the cavity mode and atomic detuning. For simplification, we choose the low atomic excitation limit, i.e., when the atoms are initially in the ground state and the average number of photons is much smaller in the excited state, so that $S_{Z} \approx <S_{Z}> \approx - N_{a}$. In addition to that, we suppose the excitation probability of a single atom to be small. In this limit, the atomic polarization can be defined in terms of the bosonic annihilation and creation operators $c = \frac{S_{-}}{\sqrt{|<S_{Z}>|}}$, which satisfy the bosonic commutation relations $[c,c^{\dagger}]=1$\cite{27}. We define $G_{a} = g \sqrt{N_{a}}$, the atom-cavity coupling strength.

\subsection{The quantum Langevin equations}
For a clear analysis of the dynamics of the system, we determine the Heisenberg equations of motion which can be obtained from the Hamiltonian \ref{hami}. Taking into consideration the effects of noise and the dissipation terms, leads to the following Heisenberg-Langevin equations 
\begin{equation}
\begin{aligned}
\dot{q_{1}} &= \omega_{m} \ p_{1}, \\
\dot{q_{2}} &= \omega_{m} \ p_{2}, \\
\dot{p_{1}} &= - \omega_{m} q_{1} - G_{0} a^{\dagger}a \cos^{2}(\theta /2) - \gamma_{m} \ p_{1} + f_{1},\\
\dot{p_{2}} &= - \omega_{m} q_{2} + G_{0} a^{\dagger}a \cos^{2}(\theta /2) - \gamma_{m} \ p_{2} + f_{2},\\
\dot{a} &= -(\kappa + i \Delta_{r})a - i G_{a} c - i G_{0}\cos^{2}(\theta /2)(q_{1} - q_{2})a + E_{L} + \sqrt{2 \kappa}a_{in},\\
\dot{c} &= -(\gamma_{a} + i \Delta_{a})c -i G_{a} a + \sqrt{2 \gamma_{a}}c_{in}.\\
\end{aligned}
\label{lang.eq}
\end{equation}
For the sake of simplicity and without loss of generality, We choose $\omega_{m_{1}} = \omega_{m_{2}} =\omega_{m}$, $\gamma_{m1} = \gamma_{m2} = \gamma_{m}$, where $\gamma_{m}$ is the mechanical damping rate and $\gamma_{a}$ is the decay rate of the atomic excited level. In equation (\ref{lang.eq}), we defined $G_{0}$ as $G_{0} = (\frac{\omega_{r}}{L})\sqrt{\frac{\hbar}{m  \omega_{m}}}$ while $f_{1}$ and $f_{2}$ are the Brownian noise operators with zero mean values. We choose the quality factor $Q_{i} = \frac{\omega_{m_{i}}}{\gamma_{m_{i}}} \gg 1$ meaning that one can assume the mechanical baths to be Markovian. Accordingly, the $f_{i} (i = 1,2)$ operator's non-zero correlation functions \cite{28,29} are given by   
\begin{equation}
\begin{aligned}
\langle f_{i}(t) f_{i}(t^{'}) + f_{i}(t^{'}) f_{i}(t) \rangle/2 \ \simeq \gamma_{m} (2 n_{th_{i}} + 1)\delta(t - t^{'}),
\end{aligned}
\end{equation}
with $n_{th_{i}} = 1/(\e^{\frac{\hbar \omega_{m_{i}}}{k_{B} t}} - 1)$ being the $i^{th}$ thermal photon number and $k_{B}$ is the Boltzmann constant. The noise operator corresponding to the atomic ensemble with zero mean value, $c_{in}$, appearing in (\ref{lang.eq}) , satisfy the non-vanishing correlations function $ \langle c_{in}(t)c_{in}^\dagger(t^{'})\rangle = \delta(t-t^{'})$ \cite{30}. 

Another kind of noise affecting the system, is the input squeezed vacuum noise operator $a_{in}$, that are fully characterized by the nonzero correlations functions \cite{31}:
\begin{equation}
\begin{aligned}
\langle \delta a_{in}^{\dagger}(t) \delta a_{in}(t^{'})\rangle &= N \delta(t - t^{'}),\\
\langle \delta a_{in}(t) \delta a_{in}^{\dagger}(t^{'}) \rangle &= (N + 1) \delta(t - t^{'}),\\
\langle \delta a_{in}(t) \delta a_{in}(t^{'})\rangle &= M \e^{-i \omega_{m} (t + t^{'})} \delta(t - t^{'}),\\
\langle \delta a_{in}^{\dagger}(t) \delta a_{in}^{\dagger}(t^{'})\rangle &= M^{*} \e^{i \omega_{m} (t + t^{'})} \delta(t - t^{'}),\\   
\end{aligned}
\end{equation}
where $M = \sinh{r}\cosh{r} e^{i \phi}$  and $N = \sinh^{2}{r}$, $r$ and $\phi$ being respectively the strength squeezing parameter and phase of the squeezed vacuum light.

\subsection{Linearization of the quantum Langevin equations}
To analyze the dynamics of the coupled system, we begin by linearizing the quantum Langevin (\ref{lang.eq}). In fact, these latter are in general nonsolvable analytically. So one needs to expand each Heisenberg operator as a sum of its classical steady state value plus an additional operator of fluctuation  with zero-mean value \cite{32}: 
\begin{equation}
\begin{aligned}
a = a_{s} + \delta a, \  q = q_{s} + \delta q, \  p = p_{s} + \delta p .
\label{xxx}
\end{aligned}
\end{equation}
The corresponding steady-state values read
\begin{equation}
\begin{aligned}
p_{1}^s &= 0,\\
p_{2}^s &= 0,\\
q_{1}^s &= \frac{- G_{0} \cos^{2}(\theta /2)|a_{s}|^2}{\omega_{m}},\\
q_{2}^s &= \frac{G_{0} \cos^{2}(\theta /2)|a_{s}|^2}{\omega_{m}},\\
c^s &= \frac{-i G_{a} a^s}{(\gamma_{a} + i \Delta_{a})},\\
a^s &= \frac{E_{L}}{\kappa + i \Delta + \frac{G_{a}^2}{\gamma_{a} + i \Delta_{a}}},\\
\end{aligned}
\end{equation}
where $\Delta = \Delta_{r} + G_{0} \cos^{2}(\theta /2)(q_{1}^s - q_{2}^s)$. 

Inserting equation (\ref{xxx}) in  (\ref{lang.eq}), and introducing $ \delta X = \frac{\delta a + \delta a^{\dagger}}{\sqrt{2}}$ ,\; $\delta Y = \frac{\delta a - \delta a^{\dagger}}{i \sqrt{2}}$ , \;  $\delta x = \frac{\delta c + \delta c^{\dagger}}{\sqrt{2}}$ , \; $\delta y = \frac{\delta c - \delta c^{\dagger}}{i \sqrt{2}} , X_{in} = \frac{a_{in} + a_{in}^\dagger}{\sqrt{2}} , \; Y_{in} = \frac{a_{in} - a_{in}^\dagger}{i \sqrt{2}} , \; x_{in} = \frac{c_{in} + c_{in}^\dagger}{\sqrt{2}} ,  \;  y_{in} = \frac{c_{in} - c_{in}^\dagger}{i \sqrt{2}}$, allows to obtain the following linearized Langevin equations:
\begin{equation}
\begin{aligned}
\delta\dot{q_{1}} &= \omega_{m} \ \delta p_{1}, \\
\delta \dot{p_{1}} &= -\omega_{m} \delta q_{1} - \gamma_{m} \ \delta p_{1}  - G \cos^{2}(\theta /2) \delta X + f_{1}, \\
\delta\dot{q_{2}} &= \omega_{m} \ \delta p_{2}, \\
\delta \dot{p_{2}} &= -\omega_{m} \delta q_{2} - \gamma_{m} \ \delta p_{2}  + G \cos^{2}(\theta /2) \delta X + f_{2}, \\
\delta \dot{X} &= - \kappa \delta X + \Delta \delta Y + G_{a} \delta y + \sqrt{2 \kappa} X_{in}, \\
\delta \dot{Y} &= -G \cos^{2}(\theta /2) \delta q_{1} + G \cos^{2}(\theta /2) \delta q_{2} - \Delta \delta X - \kappa  \delta Y - G_{a} \delta x + \sqrt{2 \kappa} Y_{in}, \\ 
\delta \dot{x} &= G_{a} \delta Y - \gamma_{a} \delta x + \Delta_{a} \delta y + \sqrt{2 \gamma_{a}} x_{in}, \\ 
\delta \dot{y} &= -G_{a} \delta X - \gamma_{a} \delta y - \Delta_{a} \delta x + \sqrt{2 \gamma_{a}} y_{in},  
\end{aligned}
\label{u}
\end{equation}
with $G = \sqrt{2} G_{0} a^s$. The resulting evolution equations of motion for the fluctuations in (\ref{u}) can be rewritten in the matrix form 
\begin{equation}
\label{diffeq}
\dot{u}(t) = A u(t) + n(t),
\end{equation}
where $u(t)$ and $n(t)$ are respectively, the column vector of the fluctuations and the column vector of noise operators, the transpose of which are respectively given by 
\begin{eqnarray}
u^{T}(\infty) &=& (\delta q_{1}(\infty),\, \delta p_{1}(\infty),\, \delta q_{2}(\infty),\, \delta p_{2}(\infty), \, \delta X(\infty), \, \delta Y(\infty), \, \delta x(\infty), \, \delta y(\infty)) \nonumber \\ 
n^T(t) &=& (0, f_{1},0, f_{2}, \sqrt{2 \kappa} X_{in}, \sqrt{2 \kappa} Y_{in}, \sqrt{2 \gamma_{a}} x_{in}, \sqrt{2 \gamma_{a}} y_{in}).
\end{eqnarray}
 The drift matrix $A$, reads
\begin{equation}
A = \begin{pmatrix} 0 & \omega_{m} & 	0 & 0 & 0 & 0 & 0 &0 \\ - \omega_{m} &  - \gamma_{m} & 0 & 0 & -G \cos^{2}(\theta /2) & 0 & 0 & 0\\ 0 & 0 & 0 & \omega_{m} & 0 & 0 & 0 & 0 \\ 0 & 0 & - \omega_{m} &  - \gamma_{m} & G \cos^{2}(\theta /2) & 0 & 0 & 0 \\ 0 & 0 & 0 & 0 & -\kappa & \Delta & 0 & G_{a} \\ -G \cos^{2}(\theta /2) & 0 & G \cos^{2}(\theta /2) & 0 & -\Delta & -\kappa & - G_{a} & 0 \\ 0 & 0 & 0 & 0 & 0 & G_{a} & -\gamma_{a} & \Delta_{a} \\ 0 & 0 & 0 & 0 & -G_{a} & 0 & -\Delta_{a} & -\gamma_{a} .
\end{pmatrix}
\end{equation}
The solution of the differential equation (\ref{diffeq}), is $u(t) = Y(t)u(0) + \int_{0}^{t} dx Y(x) \eta(t - x)  $, with $Y(t) = \exp{At}$.

\subsection{Covariance Matrix}
The covariance matrix V of the system can be obtained using the following Lyapunov equation \cite{33,34}:
\begin{equation}
\begin{aligned}
A V + V A^T = -D,
\label{18}
\end{aligned}
\end{equation}
where D is a diagonal matrix that represents the noise correlations. It is given by
\begin{equation}
D = \begin{pmatrix} 0 & 0 & 0 & 0 & 0 & 0 & 0 & 0 \\ 0 & \gamma_{m} (2 n_{th} + 1) & 0 & 0 & 0 & 0 & 0 & 0 \\ 0 & 0 & 0 & 0 & 0 & 0 & 0 & 0 \\ 0 & 0 & 0 & \gamma_{m} (2 n_{th} + 1) & 0 & 0 & 0 & 0 \\ 0 & 0 & 0 & 0 & 2 \kappa (Re[M] + N + \frac{1}{2}) & 2 \kappa Im[M] & 0 & 0 \\ 0 & 0 & 0 & 0 & 2 \kappa Im[M] & 2 \kappa (-Re[M] + N + \frac{1}{2}) & 0 & 0 \\ 0 & 0 & 0 & 0 & 0 & 0 & \gamma_{a} & 0 \\ 0 & 0 & 0 & 0 & 0 & 0 & 0 & \gamma_{a} \\
\end{pmatrix}
\end{equation}
The Covariance Matrix V  can be written in a block form:
\begin{equation}
V = \begin{pmatrix} V_{m_{1}} & V_{m_{1}m_{2}} & V_{m_{1}op} & V_{m_{1}a} \\ V_{m_{1}m_{2}}^T &  V_{m_{2}} & V_{m_{2}op}  & V_{m_{2}a}  \\ V_{m_{1}op} ^T & V_{m_{2}op}^T & V_{op} & V_{op a}  \\ V_{m_{1} a} ^T & V_{m_{2}a} ^T & V_{op a} ^T &  V_{a} 
\end{pmatrix},
\label{cov}
\end{equation}
where $V_{a}$, $V_{m_{j}}(j = 1,2)$, and $V_{op}$ are the Covariance Matrices of the atomic mode, the $(j = 1,2)$ mechanical mode and the optical mode, respectively.

When interested in studying the behavior of only two subsystems (and their correlations, among other things), the  global  $8 \times 8$ covariance matrix $V$  (\ref{cov}) can be reduced to a $4 \times 4$ submatrix $V_{S}$, containing only the covariance matrices of the subsystems of interest:
\begin{equation}
V_{S} = \begin{pmatrix} A & C  \\ C^T & B
\end{pmatrix},
\label{sub}
\end{equation}
with $A$ and $B$ being the $2 \times 2$ covariance matrices  describing the single modes and $C$ the $2 \times 2$ covariance matrix of the quantum correlations between the two subsystems.

\section{Entanglement analysis}

The system being comprised of four modes: atomic (\textit{a}), optical (\textit{op}) and the two mechanical modes ($m_1$ and $m_2$), it allows the study of various types of entanglements. Bipartite entanglement, which is generally the type that is thoroughly studied in the literature, can be investigated in this case, using any bi-partition of the system. Moreover, tripartite entanglement can also be discussed in detail using the different tri-partitions of the system. The benefit from such a general study is that the comparison of the different types of entanglement brings more insight into their general behavior and their mutual influence.

\subsection{Bipartite entanglement}
To quantify the bipartite stationary entanglement between any two modes $x$ and $y$ ($x,y = a, m_1, m_2$ or $op$), we use the logarithmic negativity $E_{N}$. It is defined for Gaussian continuous variable systems as \cite{35,36}
\begin{equation}
\begin{aligned}
E_{N} = \max [0, -\ln{ 2\tilde{\eta}}] ,
\end{aligned}
\end{equation}
with 
\begin{equation}
\begin{aligned}
\tilde{\eta} = \min\left\{  \hbox{eig} \ | \oplus_{j=1}^2 (-\sigma_y) P V_{S} P |\right\},
\end{aligned}
\end{equation}
where $\sigma_y$ is the y-Pauli matrix, $V_{S}$ is the $4 \times 4$ covariance matrix of the two subsystems and $P = \sigma_z \oplus 1$, with $\sigma_z$ being the z-Pauli matrix.

In order to evaluate numerically the logarithmic negativity. we choose, the power of the driven laser $P = 35 mW$, the masses and the frequencies of two oscillators are respectively $m = 10 \ ng$ and $\omega_{m} = 2 \pi \times 10^7 H_z$,  the laser wavelength is $\lambda = 1064 \ nm$, $\theta = \frac{\pi}{3} $,  the cavity decay rate $\kappa = \pi \times 10^7 H_z$, the mechanical damping rate $\gamma_{m} = 2 \pi \times 10^2 H_z$, the length of the cavity $l = 1 mm $, the phase of the squeezed vacuum light $\phi = 0$. We choose the parameter of the atoms to be $G_{a} = 12 \pi \times 10^6 H_z$ and $\gamma_{a} = \pi \times 10^7 H_z$. In addition, we consider that the atoms are resonant: $\Delta_{a} = -\omega_{m} $. Some parameters are taken from the set of experiments \cite{37,38,39}.

It is important to note that, the symmetry between the mechanical modes reflects in the expressions of the logarithmic negativities involving these modes. For instance the logarithmic negativity between any mechanical mode and the optical mode are the same: $E_{m1op}=E_{m2op}$, and the same with respect to the atomic mode: $E_{m1a}=E_{m2a}$.  Next, we investigate the stationary entanglement as a function of the thermal bath’s temperature $T$.

\begin{figure}[H]
\centering
\includegraphics[height=9cm]{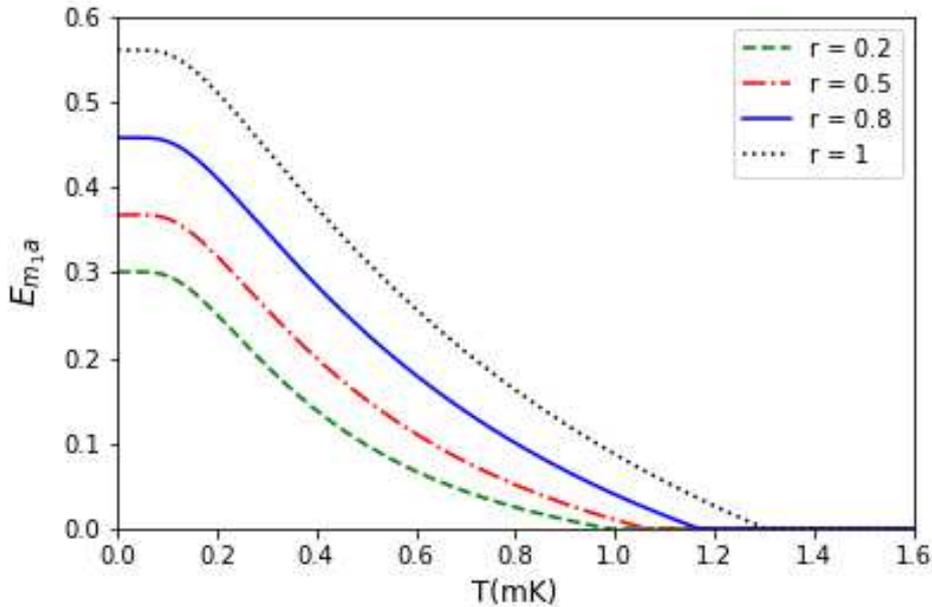}
\caption{Plot of the bipartite entanglement $E_{m_1a}$, between $m_1$ and $a$,  versus the thermal bath temperature $T(K)$, for different values of the input field squeezing parameter $r$.}
\label{EN3:2}
\end{figure}

Figure \ref{EN3:2} shows the bipartite stationary entanglement $E_{m_1a}$, we choose $\Delta = \omega_{m}$ and numerically simulate the logarithmic negativity between the mechanical mode 1 and the atomic mode for different values of the squeezing parameter $r$, since the purpose of this paper is to investigate the impact of the atomic medium on hybrid optomechanical systems. It is worth noting, that the atoms are indirectly coupled to the mechanical resonators through their common interaction with the input field. The motion of the resonator under the influence of the optical field does enhance the radiation pressure force, which allows for an atom-mirror entanglement. We see from the figure that $E_{m_1a}$  decreases as the temperature increase. It is evident that the entanglement decreases with the effect of the environment's temperature due to the thermal fluctuations. The presence of the atomic medium in an optomechanical cavity though, has a favorable effect since it enhances the optomechanical coupling. This is explained by the plateau observed for the entanglement at low temperatures. It is a generally found that adding an atomic ensemble to a cavity allows for a stronger optomechanical coupling. Indeed, the atom-field coupling strength and the excitation number have an important influence on the atomic effective damping rate of the mechanical resonators \cite{40}. 

The numerical simulation results show that the progressive injection of the squeezed light increases the entanglement, and it becomes more robust against the environment temperature, \textit{e.g.} for low temperatures, when $r = 0.2$, the numerical values of the entanglement are $E_{m_1a}  = 0.3$, whereas,  for $r = 1$, it is $E_{m_1a}  = 0.56$. Notice that, the entanglement in the case of $r = 1$ persists against the bath environment more than in the case where the parameter of squeezing $r = 0.2$. This explains the relationship between the squeezed vacuum light and entanglement due to the light-matter interaction. Indeed, the increase of the photon number in the cavity leads to a stronger radiation pressure force and does enhance the quantum correlations transfer from squeezed light to the subsystems.

\begin{figure}[H]
\centering
\includegraphics[height=9cm]{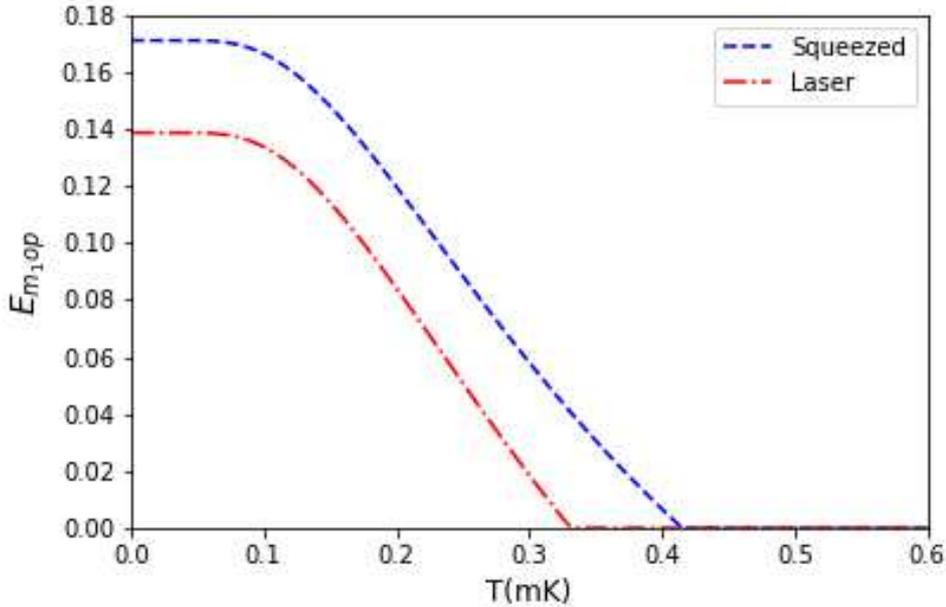}
\caption{The logarithmic negativity $E_{m_1op}$, between $m_1$ and $op$ versus the thermal bath temperature $T(K)$.The other parameters are chosen as \ref{EN3:2}.}
\label{EN2}
\end{figure}

In Figure \ref{EN2}, we plot the logarithmic negativity $E_{m_1  op}$ which express the entanglement between the mechanical mode 1 and the optical mode. We analyze the influence of the injection of the squeezing light on the hybrid system. The red line representing the entanglement when there is no squeezing light source and only the laser field is injected $(r = 0)$, shows that the entanglement vanishes around $T \simeq 0.35 mk$. However, the blue line representing the squeezed vacuum light with $(r = 0.1)$, shows that the entanglement survives until $T \simeq 0.42 mk$. We remark that for low temperatures we have $E_{m_2  op} = 0.14$ (when $r = 0$) and $E_{m_2 op} = 0.17$ (when $ r = 0.1$). A relatively significant entanglement is reached for sufficiently large number of photons from the two sources, as they allow for a robust photon-phonon interaction via the radiation pressure. Indeed, the photons exert a small push on the surface of the movable mirror and changes the cavity’s length, which in turn, modifies the intensity of the field and do enhance the radiation pressure force, and allows for a strong optomechanical coupling that optimizes entanglement \cite{41}. In addition to that, we have the quantum fluctuations from the optical environment and the thermal fluctuations from the mechanical bath environment inducing decoherence. Since any genuine feasible system couples with its own environment in some way, a large number of photons is needed to circumvent the decohering effects of the quantum fluctuations and strengthens the resulting entanglement.

\subsection{Tripartite entanglement}

In order to study the existence of genuine tripartite entanglement we adopt a quantitative measure of tripartite negativity  \cite{0042,42,0142}, which for a tripartite system $(ABC)$ is given by
\begin{equation}
\label{LNtripartite}
\begin{aligned}
{\cal{E}}_{ABC} = (E_{A|BC} \ E_{B|AC} \ E_{C|AB})^{1/3}.
\end{aligned}
\end{equation}
$E_{A|BC} = \max [0 , -\ln 2 \nu_{A|BC}]$ is the logarithmic negativity of the one mode-versus-two modes bipartitions in the system, where $\nu_{A|BC} = \min\left\{  \hbox{eig} \ | \oplus_{j=1}^3 (-\sigma_y) P_{A|BC} \ V_{3} \ P_{A|BC} |\right\}$, with $P_{A|BC} = \sigma_z \oplus 1 \oplus 1$, $P_{B|AC} = 1 \oplus \sigma_z \oplus  1$ and $P_{C|AB} = 1 \oplus 1 \oplus \sigma_z $ are the matrices of the partial transposition of the tripartite covariance matrix, and $V_{3}$ is the $6 \times 6$ covariance matrix of the tripartite system.
In our case $A$, $B$ and $C$ can be one of the mechanical modes, optical mode or atomic mode. Moreover, for the sake of simplicity we will use the following compact notations  ${\cal{E}}_{1}\equiv {\cal{E}}_{m_1m_2a}$, ${\cal{E}}_{2}\equiv {\cal{E}}_{a\,m_1op}={\cal{E}}_{a\,m_2op}$ and ${\cal{E}}_{3}\equiv {\cal{E}}_{m_1m_2op}$.

\begin{figure}
\subfloat[]{\begin{minipage}[c][1\width]{0.5\textwidth}
\centering
\includegraphics[height=7cm, width=1.2\textwidth]{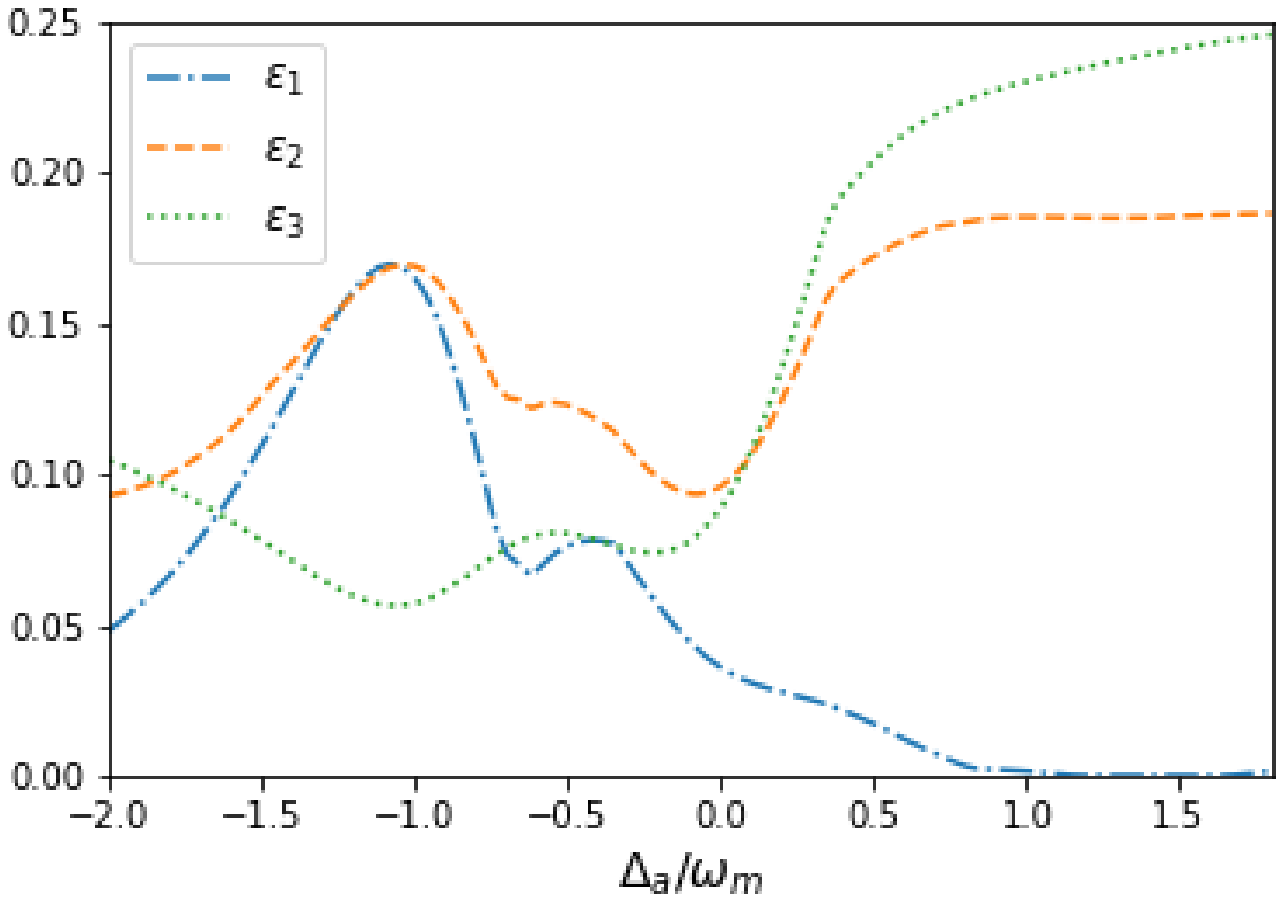}
\label{fig4tri}
\end{minipage}}
\hspace{1cm}
\hfill
\subfloat[]{\begin{minipage}[c][1\width]{0.5\textwidth}
\centering
\includegraphics[height=7cm,width=1.2\textwidth]{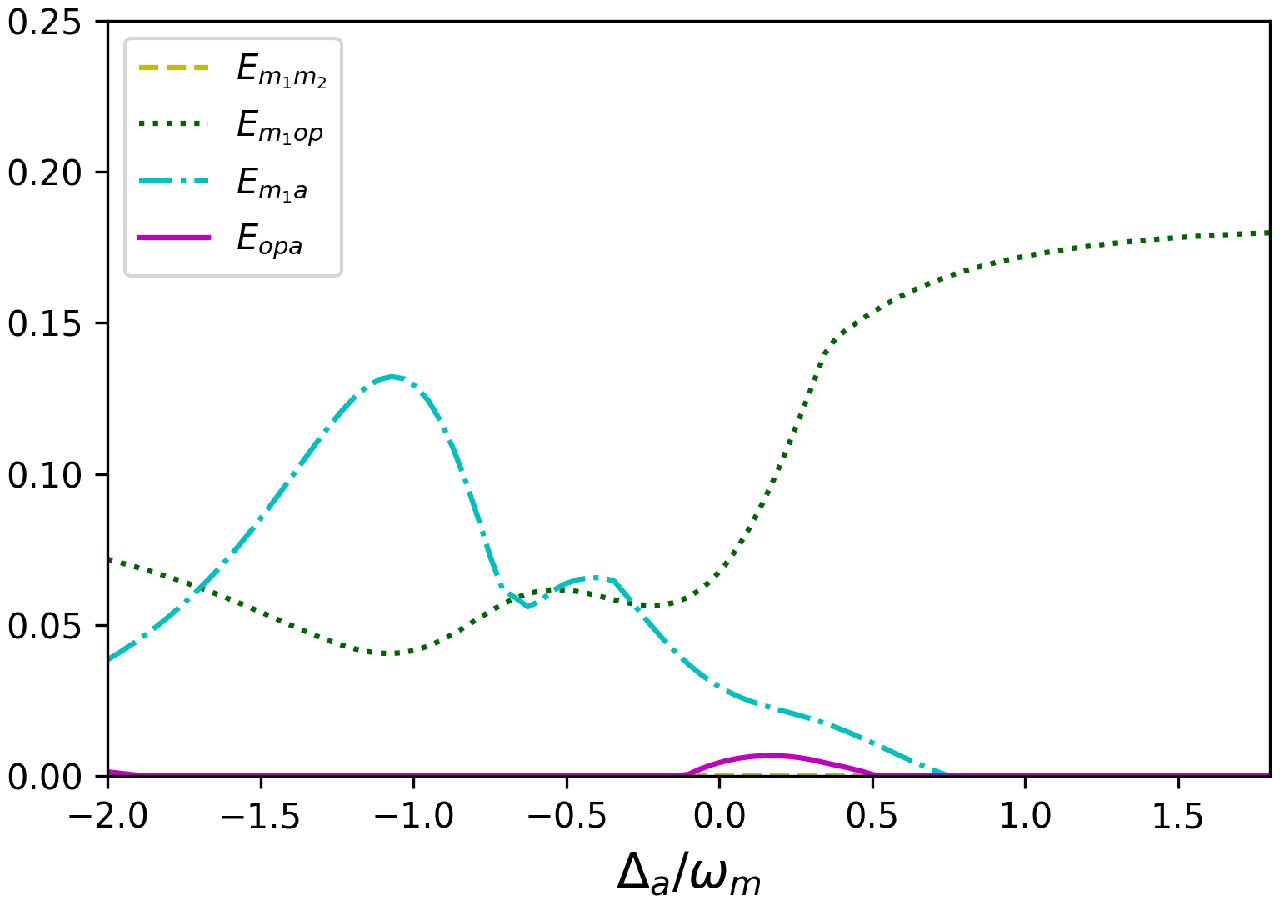}
\label{fig4bi}
\end{minipage}}
\caption{Effect of the normalized atomic detuning on the (a) tripartite logarithmic negativities  ${\cal{E}}_{1}$, ${\cal{E}}_{2}$ and ${\cal{E}}_{3}$ and (b) bipartite negativities ($E_{m_1 m_2}$, $E_{m_1 op}$, $E_{m_1 a}$ and $E_{a \, op}$)  . In all the cases we choose $\Delta = \omega_{m}$, $P = 10 mW$, $r = 0.1$ and the temperature $T = 0.1 mK$.}
\label{fig4}
\end{figure}

We plot in figure \ref{fig4tri}, the tripartite entanglement as captured by the logarithmic negativity (\ref{LNtripartite}) versus the normalized atomic detuning. The three logarithmic negativities react quite differently, and broadly speaking, it depends on whether the atomic mode is involved or not. For instance,  ${\cal{E}}_{1}$ and ${\cal{E}}_{2}$ have their maximum close to the region where $\Delta_{a} = - \omega_{m}$ whereas, ${\cal{E}}_{3}$ decreases in this interval and actually reaches its minimum in this region. It is remarkable that the tripartite entanglement between the optical mode, the atomic mode and the mechanical mode (${\cal{E}}_{2}$ ), is the one that remains significant in a broader interval.
This is due to the absorption and remission of photons by the atoms. On the other hand, the more photons we have in the cavity, the more the photon-phonon interaction is enhanced via the radiation pressure force due to the effect of the vibrating mirror. Under the atom-photon-phonon interaction, not only the region of the effective detuning is wider, but also a significant entanglement is obtained. These collective bosonic modes form an optimal quantum tripartite system that ensures a strong entanglement sharing.

It is worth noticing that, when $\Delta_{a} > 0 $, both ${\cal{E}}_{2}$ and ${\cal{E}}_{3}$ asymptotically increase, while ${\cal{E}}_{1}$ is negligible
i.e, the tripartite $\left\{m_1,m_2,a\right\}$ entanglement is not present in this area. This result might be interpreted as a result of the negative effective atomic detuning being a convenient choice since it regulates the evolution of the atomic quadrature \cite{43}.

This collective tripartite behavior can be explained by the corresponding underlying bipartite entanglement shown in Figure \ref{fig4bi}. As a matter of fact, the behavior of ${\cal{E}}_{1}\equiv {\cal{E}}_{m1m2a}$ in Figure \ref{fig4tri} is identical to that of $E_{{m_1}a}$ in Figure \ref{fig4bi} because the other bipartite entanglement $E_{{m_1}{m_2}}$ involved in ${\cal{E}}_{1}$ is shown in Figure \ref{fig4bi} to be negligible. Similarly, ${\cal{E}}_{3}\equiv {\cal{E}}_{m1m2op}$, relies on the \textit{mechanical-optical}, entanglement $E_{{m_1}{op}}$ and the negligible \textit{mechanical-mechanical}, $E_{{m_1}{m_2}}$ entanglement. This results in the behavior of ${\cal{E}}_{3}$ resembling that of $E_{{m_1}{op}}$. In contrast, ${\cal{E}}_{2}\equiv {\cal{E}}_{m1 a\, op}$, depends on three bipartite entanglements $E_{{m_1}op}$, $E_{{m_1}a}$ and $E_{op\, a}$, the last one of which is shown to be negligible in Figure \ref{fig4bi}. ${\cal{E}}_{2}$ is thus mainly driven by $E_{{m_1}{op}}$ and $E_{{m_1}{a}}$; henceforth, the behavior of ${\cal{E}}_{2}$ in Figure \ref{fig4tri} is a hybrid of that of $E_{{m_1}{op}}$ and $E_{{m_1}{a}}$ in Figure \ref{fig4bi}.

This discussion confirms the premise that the \textit{mechanical}-\textit{optical}-\textit{atomic} modes forms an ideal system, which requires that $\Delta_{a} = - \omega_{m}$, thus providing optimal performances on quantum information theory of continuous variable systems.

\begin{figure}[h]
\subfloat[]{\begin{minipage}[c][1\width]{0.5\textwidth}
\centering
\includegraphics[height=7cm, width=1.2\textwidth]{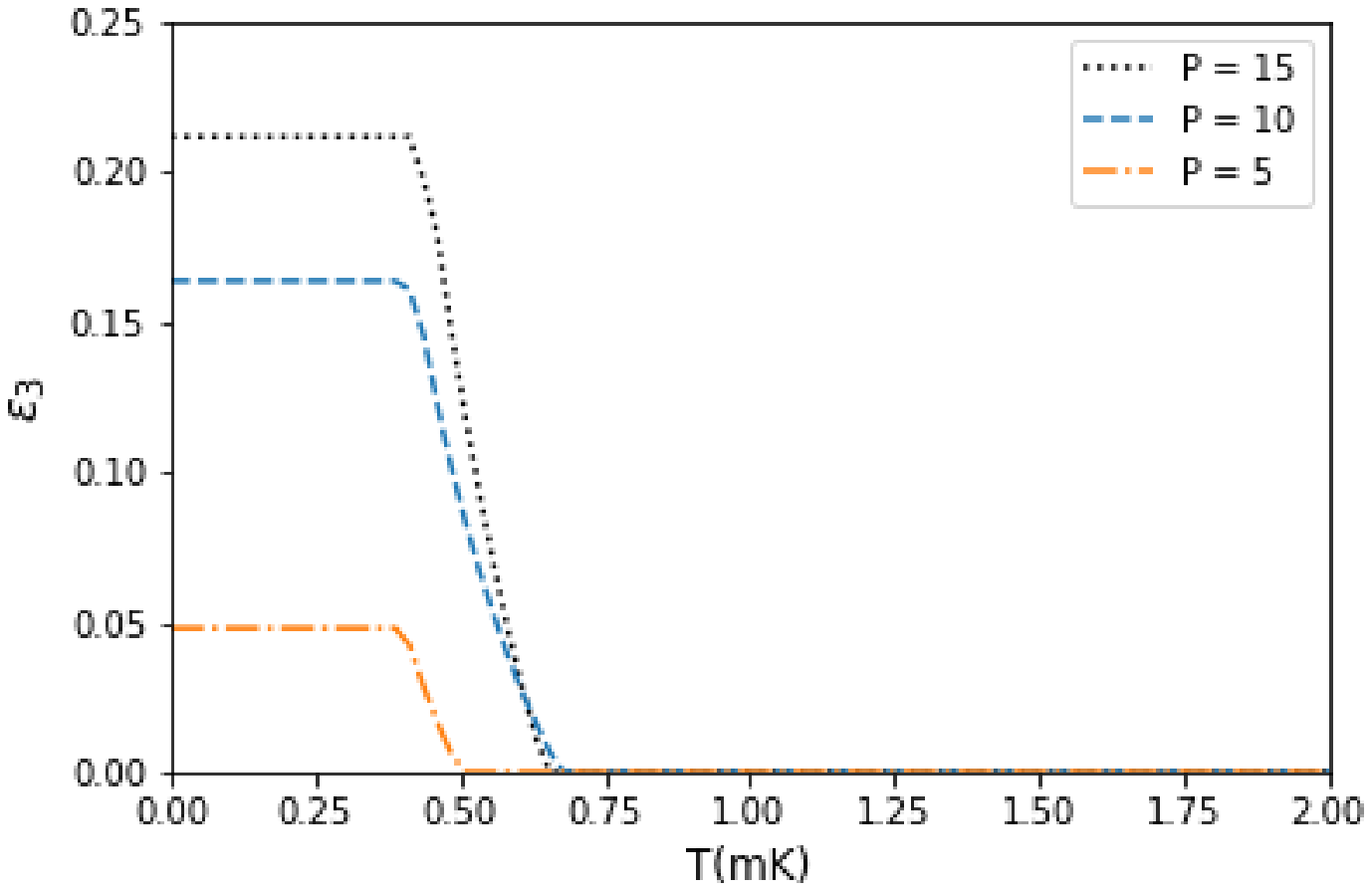}
\end{minipage}}
\hspace{1cm}
\hfill
\subfloat[]{\begin{minipage}[c][1\width]{0.5\textwidth}
\centering
\includegraphics[height=7cm,width=1.2\textwidth]{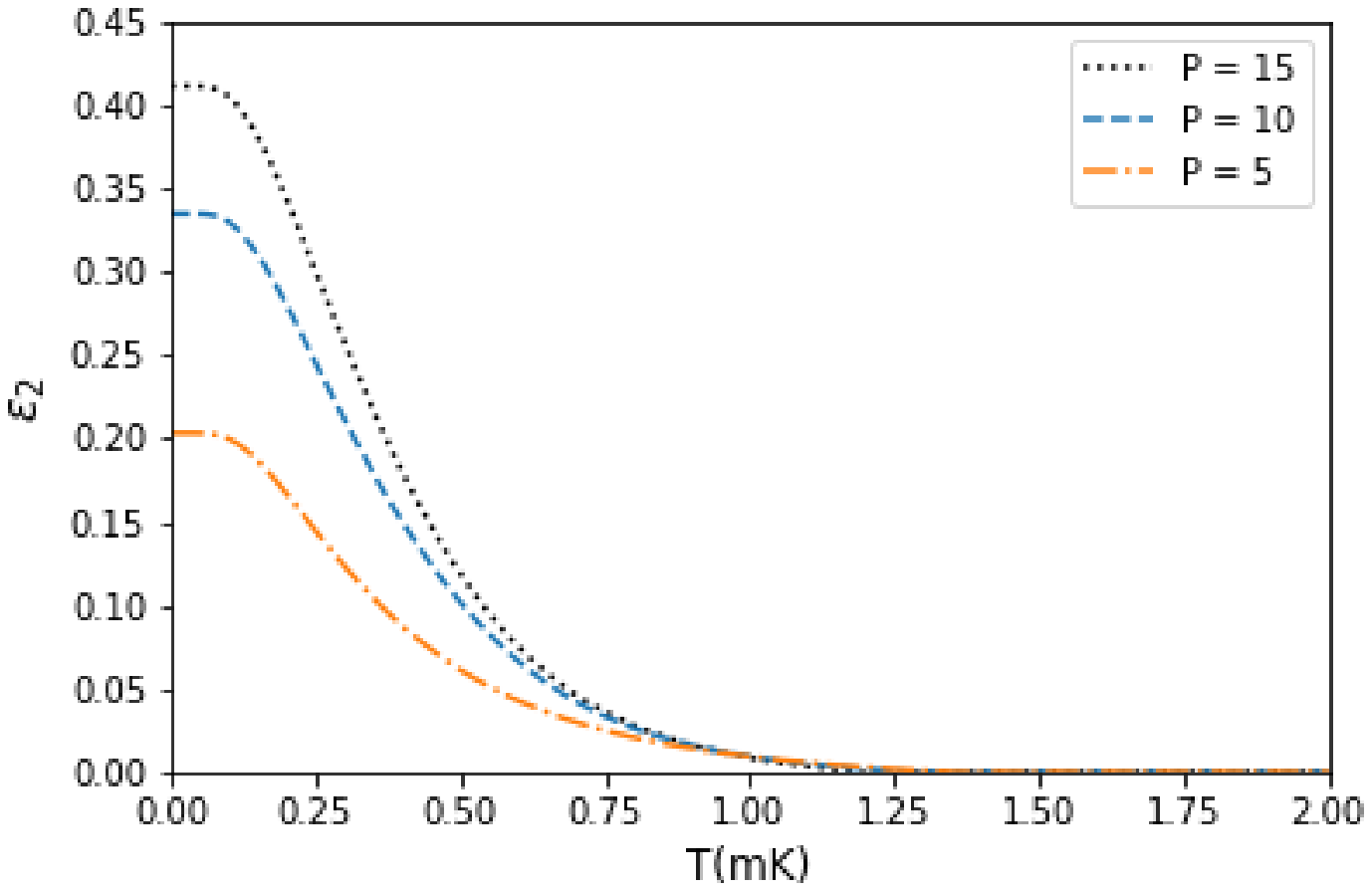}
\end{minipage}}
\caption{(a) ${\cal{E}}_{2}$ and (b) ${\cal{E}}_{3}$ as a function of the temperature $T(mK)$ for different values of the pumping power $P$ (measured in mW). $\Delta = 0.5 \omega_{m}$ and the other parameters are chosen as in figure \ref{fig4}.}
\label{figure5}
\end{figure}

To analyze the influence of the pumping power, we plotted in figure \ref{figure5} ${\cal{E}}_{2}$ and ${\cal{E}}_{3}$ as a function of the environment temperature $T$, for different values of $P$. With the increase of driving power, the values of logarithmic negativity increase and, relatively, resist better the environment-induced decoherence. We also find that, for $P =15 mW$, ${\cal{E}}_{3}$ vanishes at $T \simeq 0.65 mk$, while ${\cal{E}}_{2}$ survives until $T \simeq 1.2mk$, so ${\cal{E}}_{3}$ vanishes quicker than ${\cal{E}}_{2}$ due to the thermal environment. On the other hand, for $P = 5 mW$ and for a fixed temperature $T= 0.27 mk$, ${\cal{E}}_{3} = 0.048$ while ${\cal{E}}_{2} = 0.13$, which shows that, compared to ${\cal{E}}_{2}$, ${\cal{E}}_{3}$ requires much higher values of power that increases the number of photons in the cavity and a very low thermal phonons numbers, i.e. temperature. It is an intuitive fact that entanglement is highly sensitive to the  fluctuations, with the mechanical modes being much noisier and the fluctuations of the atomic mode being less noisy resulting from the collision between photons of the light source and the photons emitted by the atoms. This partly explains the reason of the fragility of ${\cal{E}}_{3}$ compared to ${\cal{E}}_{2}$  as this latter depends less on the much more fragile correlations involving the mechanical modes than the former. In addition to that, we have an additional squeezed vacuum light, the injection of which reduces the fluctuations and results in stronger correlations when it is involved; this illustrates well the usefulness of squeezed light.
The system under consideration is an optical ring cavity, which is characterized by a high quality factor since the use of two movable mirrors enhances the field-mirrors interaction. With significant values of the power, increases the circulation inside the cavity which in turns enhances the interaction and thus the entanglement, establishing the importance of the parameter of power as well.

\begin{figure}[h]
\subfloat[]{\begin{minipage}[c][1\width]{0.5\textwidth}
\centering
\includegraphics[height=7cm, width=1.2\textwidth]{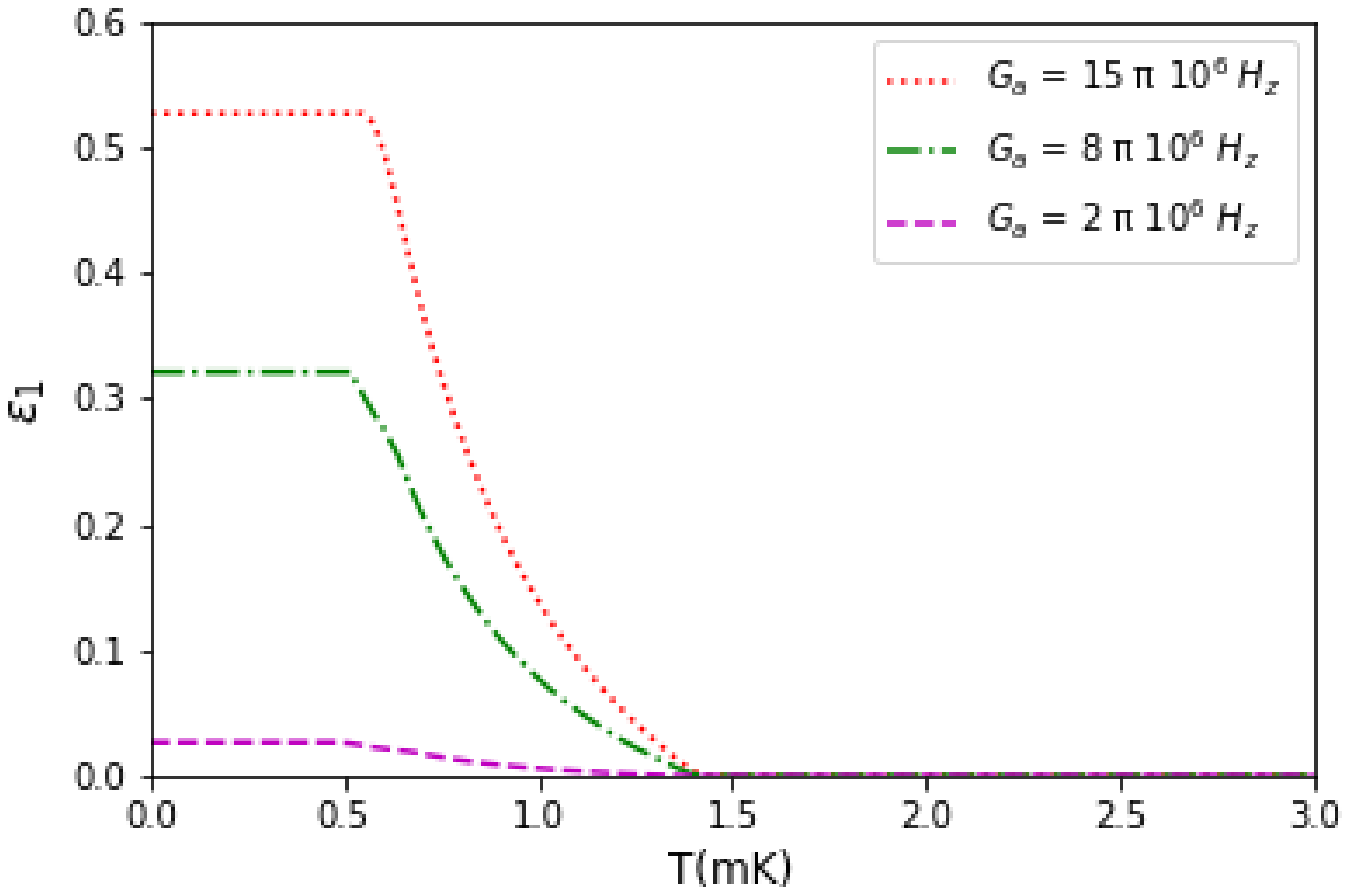}
\label{G01}
\end{minipage}}
\hspace{1cm}
\hfill
\subfloat[]{\begin{minipage}[c][1\width]{0.5\textwidth}
\centering
\includegraphics[height=7cm,width=1.2\textwidth]{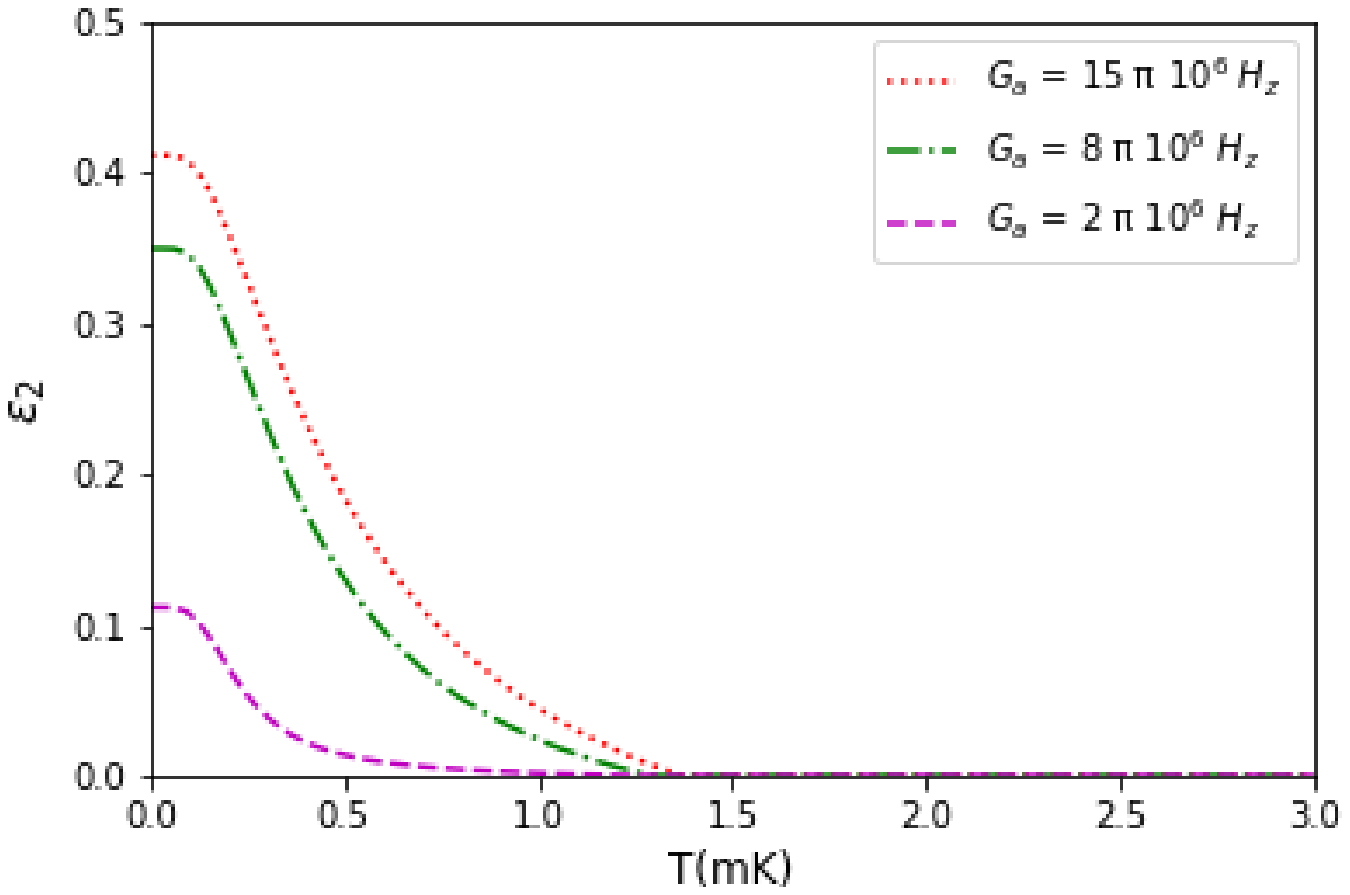}
\label{G02}
\end{minipage}}
\caption{Plot of the influence of the atom cavity coupling $G_{a}$ on ${\cal{E}}_{1}$ and ${\cal{E}}_{2}$ as a function of the temperature $T(mK)$. We choose $P = 35 mW$, $r = 0.6$, $\Delta = \omega_{m}$.}
\end{figure}

Let us finally explore the effect of the atomic-field coupling strength $G_{a}$. To that, we simulate the tripartite entanglement ${\cal{E}}_{1}$ and ${\cal{E}}_{2}$ versus the temperature for different values of $G_{a}$. It is observed that, for a fixed temperature \textit{e.g.} $T = 0.5 mK$ as $G_{a}$ increases, both ${\cal{E}}_{1}$ and ${\cal{E}}_{2}$ increase and in general, one can achieve significant amounts of tripartite entanglement for high couplings. This can, for instance, be achieved by increasing the atom numbers in the cavity and with a strong driven laser that increases the number of photons as explained in previously, then a well-fortified interaction is achieved leading to a strong atom-field coupling. Indeed, for high-quality cavities such as ring cavity considerable coupling can be easily realized experimentally \cite{44}. By comparing the plots \ref{G01} and \ref{G02}, it comes out that ${\cal{E}}_{1}$ is more resistant than  ${\cal{E}}_{2}$ to the thermal induced decoherence; \textit{e.g.}, when $G_{a} = 8 \pi 10^6 H_{z}$, ${\cal{E}}_{2}$ vanishes at $T \approx 1.3 mk$ while ${\cal{E}}_{1}$  vanishes at $T \approx 1.5 mk$ all the while maintaining a significant amount and plateauing over a longer interval. It is clearly seen that ${\cal{E}}_{1}$  \textit{i.e.} the $\left\{m_1,m_2,a\right\}$ entanglement which is of interest in this paper is greater than ${\cal{E}}_{2}$ \textit{i.e.}  $\left\{a,m_1,op\right\}$ even though atoms and mirrors are only indirectly coupled in the Hamiltonian \ref{hami}. As a matter of fact, ${\cal{E}}_{1}$ enhances at the expanse of the field-mirror interaction via the radiation pressure as in \cite{23}. According to all these results, we conclude that a significant entanglement is established with a strong atom-field coupling and can be differentiated according to the optical field squeezing.

\section{conclusion}
In this work, we have proposed a theoretical scheme for the study of steady state bipartite and tripartite entanglement. The system under study is a ring cavity with fixed mirror and two movable ones, an atomic ensemble of two-level atoms confined in the cavity. This latter is pumped with a coherent laser source and a squeezed vacuum light driving to enhance the quantum correlations. A proper analysis of the dynamics of the coupled system is made allowing to get the set of the quantum Langevin equations and the linearization of the equations is carried to get the $8 \times 8$ steady state covariance matrix that is fully describing the hybrid optomechanical system (cavity field – atomic ensemble - two vibrating mirrors). The bipartite and tripartite logarithmic negativity are used to evaluate the entanglement of the multimode system.

An analysis of the bipartite and tripartite entanglement being possibly shared by the coupled system confirms the negative influence of the environment’s temperature on the different types of entanglement. Some being more resilient than other types; in particular the benefits of adding the atomic ensemble in the cavity is observed as it allows for more resistant entanglement. In this regard, we have shown that a stronger atomic-field coupling allows for better entanglement. On the other hand, with the increase of the power of the driving laser, significant entanglement and broader effective detuning region can be achieved. In addition to that, we have shown the needfulness of the squeezed light source since this latter allows for a strong entanglement.

Such a scheme will open new perspectives for the application of quantum teleportation in the cavity and the implementation of quantum memories for continuous variable quantum information processing. The enhancement of the entanglement in cavity, will prove critical for the cavity optomechanical sensing \cite{persp_sens,persp-daniel}. The presence of a strong multipartite entanglement can be exploited for realizing teleportation in optomechanical ring cavities in the same spirit of what was achieved in \cite{perspective}.

\bibliography{bibliography.bib}

\begin{thebibliography}{10}

\bibitem{1}
Carlton~M Caves.
\newblock Quantum-mechanical radiation-pressure fluctuations in an
  interferometer.
\newblock {\em Physical Review Letters}, 45(2):75, 1980.

\bibitem{2}
Thomas Corbitt, David Ottaway, Edith Innerhofer, Jason Pelc, and Nergis
  Mavalvala.
\newblock Measurement of radiation-pressure-induced optomechanical dynamics in
  a suspended fabry-perot cavity.
\newblock {\em Physical Review A}, 74(2):021802, 2006.

\bibitem{3}
David Vitali, Sylvain Gigan, Anderson Ferreira, HR~B{\"o}hm, Paolo Tombesi,
  Ariel Guerreiro, Vlatko Vedral, Anton Zeilinger, and Markus Aspelmeyer.
\newblock Optomechanical entanglement between a movable mirror and a cavity
  field.
\newblock {\em Physical review letters}, 98(3):030405, 2007.

\bibitem{4}
XuBo Zou and W~Mathis.
\newblock Quantum information processing and entanglement with josephson charge
  qubits coupled through nanomechanical resonator.
\newblock {\em Physics Letters A}, 324(5-6):484--488, 2004.

\bibitem{5}
Tobias~J Kippenberg and Kerry~J Vahala.
\newblock Cavity opto-mechanics.
\newblock {\em Optics express}, 15(25):17172--17205, 2007.

\bibitem{6}
Claude Fabre, Michel Pinard, Sophie Bourzeix, Antoine Heidmann, Elisabeth
  Giacobino, and Serge Reynaud.
\newblock Quantum-noise reduction using a cavity with a movable mirror.
\newblock {\em Physical Review A}, 49(2):1337, 1994.

\bibitem{7}
Jing Zhang, Kunchi Peng, and Samuel~L Braunstein.
\newblock Quantum-state transfer from light to macroscopic oscillators.
\newblock {\em Physical Review A}, 68(1):013808, 2003.

\bibitem{8}
Junhong Li, Bangpin Hou, Yonghong Zhao, and Lianfu Wei.
\newblock Enhanced entanglement between two movable mirrors in an
  optomechanical system with nonlinear media.
\newblock {\em EPL (Europhysics Letters)}, 110(6):64004, 2015.

\bibitem{9}
AD~Armour, MP~Blencowe, and KC~Schwab.
\newblock Quantum dynamics of a cooper-pair box coupled to a micromechanical
  resonator.
\newblock {\em Phys Rev Lett}, 88:148301, 2002.

\bibitem{10}
David Vitali, Stefano Mancini, Luciano Ribichini, and Paolo Tombesi.
\newblock Macroscopic mechanical oscillators at the quantum limit through
  optomechanical cooling.
\newblock {\em JOSA B}, 20(5):1054--1065, 2003.

\bibitem{11}
Philipp Treutlein, David Hunger, Stephan Camerer, Theodor~W H{\"a}nsch, and
  Jakob Reichel.
\newblock Bose-einstein condensate coupled to a nanomechanical resonator on an
  atom chip.
\newblock {\em Physical review letters}, 99(14):140403, 2007.

\bibitem{12}
C~Genes, D~Vitali, and P~Tombesi.
\newblock Emergence of atom-light-mirror entanglement inside an optical cavity.
\newblock {\em Physical Review A}, 77(5):050307, 2008.

\bibitem{13}
H~Ian, ZR~Gong, Yu-xi Liu, CP~Sun, and Franco Nori.
\newblock Cavity optomechanical coupling assisted by an atomic gas.
\newblock {\em Physical Review A}, 78(1):013824, 2008.

\bibitem{14}
Klemens Hammerer, Markus Aspelmeyer, Eugene~Simon Polzik, and Peter Zoller.
\newblock Establishing einstein-poldosky-rosen channels between nanomechanics
  and atomic ensembles.
\newblock {\em Physical review letters}, 102(2):020501, 2009.

\bibitem{15}
David Hunger, Stephan Camerer, Theodor~W H{\"a}nsch, Daniel K{\"o}nig,
  J{\"o}rg~P Kotthaus, Jakob Reichel, and Philipp Treutlein.
\newblock Resonant coupling of a bose-einstein condensate to a micromechanical
  oscillator.
\newblock {\em Physical Review Letters}, 104(14):143002, 2010.

\bibitem{16}
Rina Kanamoto and Pierre Meystre.
\newblock Optomechanics of ultracold atomic gases.
\newblock {\em Physica Scripta}, 82(3):038111, 2010.

\bibitem{18}
Subhadeep Chakraborty and Amarendra~K Sarma.
\newblock Entanglement dynamics of two coupled mechanical oscillators in
  modulated optomechanics.
\newblock {\em Physical Review A}, 97(2):022336, 2018.

\bibitem{17}
Subhadeep Chakraborty and Amarendra~K Sarma.
\newblock Qubit assisted enhancement of quantum correlations in an
  optomechanical system.
\newblock {\em Annals of Physics}, 392:39--48, 2018.

\bibitem{19}
Yong-Chun Liu, Yu-Wen Hu, Chee~Wei Wong, and Yun-Feng Xiao.
\newblock Review of cavity optomechanical cooling.
\newblock {\em Chinese Physics B}, 22(11):114213, 2013.

\bibitem{20}
C~Genes, D~Vitali, and P~Tombesi.
\newblock Emergence of atom-light-mirror entanglement inside an optical cavity.
\newblock {\em Physical Review A}, 77(5):050307, 2008.

\bibitem{21}
Cheng-Hua Bai, Dong-Yang Wang, Hong-Fu Wang, Ai-Dong Zhu, and Shou Zhang.
\newblock Robust entanglement between a movable mirror and atomic ensemble and
  entanglement transfer in coupled optomechanical system.
\newblock {\em Scientific reports}, 6(1):1--11, 2016.

\bibitem{22}
Xihua Yang, Jiaqi Liu, Xiaona Yan, and Min Xiao.
\newblock Enhanced multipartite entanglement via quantum coherence with an
  atom-assisted optomechanical system.
\newblock {\em Journal of Physics B: Atomic, Molecular and Optical Physics},
  51(20):205501, 2018.

\bibitem{23}
Sh~Barzanjeh, MH~Naderi, and M~Soltanolkotabi.
\newblock Steady-state entanglement and normal-mode splitting in an
  atom-assisted optomechanical system with intensity-dependent coupling.
\newblock {\em Physical Review A}, 84(6):063850, 2011.

\bibitem{24}
Sumei Huang and GS~Agarwal.
\newblock Entangling nanomechanical oscillators in a ring cavity by feeding
  squeezed light.
\newblock {\em New Journal of Physics}, 11(10):103044, 2009.

\bibitem{25}
Ivan~S Grudinin, Hansuek Lee, Oskar Painter, and Kerry~J Vahala.
\newblock Phonon laser action in a tunable two-level system.
\newblock {\em Physical review letters}, 104(8):083901, 2010.

\bibitem{26}
Long Chang, Xiaoshun Jiang, Shiyue Hua, Chao Yang, Jianming Wen, Liang Jiang,
  Guanyu Li, Guanzhong Wang, and Min Xiao.
\newblock Parity--time symmetry and variable optical isolation in
  active--passive-coupled microresonators.
\newblock {\em Nature photonics}, 8(7):524--529, 2014.

\bibitem{27}
T~Holstein and Hl~Primakoff.
\newblock Field dependence of the intrinsic domain magnetization of a
  ferromagnet.
\newblock {\em Physical Review}, 58(12):1098, 1940.

\bibitem{28}
Rafael Benguria and Mark Kac.
\newblock Quantum langevin equation.
\newblock {\em Physical review letters}, 46(1):1, 1981.

\bibitem{29}
Vittorio Giovannetti and David Vitali.
\newblock Phase-noise measurement in a cavity with a movable mirror undergoing
  quantum brownian motion.
\newblock {\em Physical Review A}, 63(2):023812, 2001.

\bibitem{30}
Crispin~W Gardiner and Peter Zoller.
\newblock Quantum noise, vol. 56 of springer series in synergetics.
\newblock {\em Springer--Verlag, Berlin}, 97:98, 2000.

\bibitem{31}
CW~Gardiner.
\newblock Inhibition of atomic phase decays by squeezed light: A direct effect
  of squeezing.
\newblock {\em Physical review letters}, 56(18):1917, 1986.

\bibitem{32}
DF~Walls and GJ~Milburn.
\newblock Quantum optics springer-verlag.
\newblock {\em New York}, 1994.

\bibitem{33}
David Vitali, Sylvain Gigan, Anderson Ferreira, HR~B{\"o}hm, Paolo Tombesi,
  Ariel Guerreiro, Vlatko Vedral, Anton Zeilinger, and Markus Aspelmeyer.
\newblock Optomechanical entanglement between a movable mirror and a cavity
  field.
\newblock {\em Physical review letters}, 98(3):030405, 2007.

\bibitem{34}
DL~Elliott.
\newblock Stability theory [book reviews].
\newblock {\em IEEE Transactions on Automatic Control}, 41(3):473, 1996.

\bibitem{35}
Guifr{\'e} Vidal and Reinhard~F Werner.
\newblock Computable measure of entanglement.
\newblock {\em Physical Review A}, 65(3):032314, 2002.

\bibitem{36}
Martin~B Plenio.
\newblock Logarithmic negativity: a full entanglement monotone that is not
  convex.
\newblock {\em Physical review letters}, 95(9):090503, 2005.

\bibitem{37}
Schwab Gigan, HR~B{\"o}hm, Mauro Paternostro, Florian Blaser, G~Langer,
  JB~Hertzberg, Keith~C Schwab, Dieter B{\"a}uerle, Markus Aspelmeyer, and
  Anton Zeilinger.
\newblock Self-cooling of a micromirror by radiation pressure.
\newblock {\em Nature}, 444(7115):67--70, 2006.

\bibitem{38}
Olivier Arcizet, P-F Cohadon, T~Briant, M~Pinard, A~Heidmann, J-M Mackowski,
  Christine Michel, L~Pinard, O~Fran{\c{c}}ais, and L~Rousseau.
\newblock High-sensitivity optical monitoring of a micromechanical resonator
  with a quantum-limited optomechanical sensor.
\newblock {\em Physical review letters}, 97(13):133601, 2006.

\bibitem{39}
Thomas Corbitt, Christopher Wipf, Timothy Bodiya, David Ottaway, Daniel Sigg,
  Nicolas Smith, Stanley Whitcomb, and Nergis Mavalvala.
\newblock Optical dilution and feedback cooling of a gram-scale oscillator to
  6.9 mk.
\newblock {\em Physical Review Letters}, 99(16):160801, 2007.

\bibitem{40}
Wei Zeng, Wenjie Nie, Ling Li, and Aixi Chen.
\newblock Ground-state cooling of a mechanical oscillator in a hybrid
  optomechanical system including an atomic ensemble.
\newblock {\em Scientific reports}, 7(1):1--10, 2017.

\bibitem{41}
Markus Aspelmeyer, Tobias~J Kippenberg, and Florian Marquardt.
\newblock Cavity optomechanics.
\newblock {\em Reviews of Modern Physics}, 86(4):1391, 2014.

\bibitem{0042}
F~Anza, B~Militello, and A~Messina.
\newblock Tripartite thermal correlations in an inhomogeneous spin--star
  system.
\newblock {\em Journal of Physics B: Atomic, Molecular and Optical Physics},
  43(20):205501, 2010.

\bibitem{42}
Carlos Sab{\'\i}n and Guillermo Garc{\'\i}a-Alcaine.
\newblock A classification of entanglement in three-qubit systems.
\newblock {\em The european physical journal D}, 48(3):435--442, 2008.

\bibitem{0142}
Fabrizio Buscemi and Paolo Bordone.
\newblock Measure of tripartite entanglement in bosonic and fermionic systems.
\newblock {\em Physical Review A}, 84(2):022303, 2011.

\bibitem{43}
Gabriele De~Chiara, Mauro Paternostro, and G~Massimo Palma.
\newblock Entanglement detection in hybrid optomechanical systems.
\newblock {\em Physical Review A}, 83(5):052324, 2011.

\bibitem{44}
Rui-Jie Xiao, Gui-Xia Pan, and Ling Zhou.
\newblock Multiple optomechanically induced transparency in a ring cavity
  optomechanical system assisted by atomic media.
\newblock {\em International Journal of Theoretical Physics},
  54(10):3665--3675, 2015.

\bibitem{persp_sens}
Bei-Bei Li, Lingfeng Ou, Yuechen Lei, and Yong-Chun Liu.
\newblock Cavity optomechanical sensing.
\newblock {\em Nanophotonics}, 10(11):2799--2832, 2021.

\bibitem{persp-daniel}
Fabienne Schneiter, Sofia Qvarfort, Alessio Serafini, Andr{\'e} Xuereb, Daniel
  Braun, Dennis R{\"a}tzel, and David~Edward Bruschi.
\newblock Optimal estimation with quantum optomechanical systems in the
  nonlinear regime.
\newblock {\em Physical Review A}, 101(3):033834, 2020.

\bibitem{perspective}
Sebastian~G Hofer, Witlef Wieczorek, Markus Aspelmeyer, and Klemens Hammerer.
\newblock Quantum entanglement and teleportation in pulsed cavity
  optomechanics.
\newblock {\em Physical Review A}, 84(5):052327, 2011.

\end{thebibliography}
\bibliographystyle{unsrt}
\end{document}